\documentclass[12pt,twoside,a4paper,useAMS]{meteoroids2013}
\usepackage{color,graphicx}
\usepackage[english]{babel}

\title[Multiple maxima meteor light curves] 
{ Meteoroid structure and ablation implications from multiple maxima meteor light curves} 

\author[Roberts, Hawkes, Weryk, Campbell-Brown, Brown, Stokan and Subasinghe]   
{Roberts, I.D.$^1$, %
 Hawkes,R.L.$^1$,  
 Weryk,R.J.$^2$,  %
 Campbell-Brown,M.D.$^2$,  %
 Brown,P.G.$^{2,3}$,  %
  Stokan, E.$^2$ %
 \and Subasinghe, D.$^2$}

\affiliation{$^1$Department of Physics, Mount Allison University, Sackville, Canada (email: rhawkes@mta.ca)
\\[\affilskip]
$^2$Department of Physics and Astronomy, University of
Western Ontario, London, Canada
\\[\affilskip]
$^3$Centre for Planetary Science and Exploration, University of
Western Ontario, London, Canada}

\pagerange{110--116}
\date{Jan. 8, 2013 and in revised form Feb. 20, 2014}
\jname{Proceedings of the Meteoroids 2013 Conference\\
       Aug. 26-30, 2013, A.M. University, Pozna\'{n}, Poland}
\editors{Jopek T.J., Rietmeijer F., Watanabe J., Williams I.P., ed.}
\begin{document}
\maketitle
\begin{abstract}
The Canadian Automated Meteor Observatory (CAMO) detects occasional meteors with two maxima in the image intensified CCD based light curves. We report early results from an analysis of 21 of these events.  Most of these events show qualitatively similar light curves, with a rounded first luminous peak, followed by an almost linear sharp rise in the second peak, and a relatively rapid curved decay of the second peak. While a number of mechanisms could explain two maxima in the light curves, numerical modelling shows that most of these events can be matched by a simple dustball model in which some grains have been released well before intensive ablation begins, followed by a later release of core grains at a single time.  Best fits to observations are obtained with the core grains being larger than the pre-released outer grains, with the core grains typically $10^{-6}$~kg while the early release grains are of the order of $10^{-9}$~kg. 

\keywords{Meteoroid structure, meteor light curve, dustball, intensified CCD, fragmentation. }
\end{abstract}
\section{Introduction}
Light curves provide one of the best indicators of the mode of meteor  ablation and fragmentation, and implied meteoroid structure.  The intensity of the light produced is indicative of the almost instantaneous mass loss rate.  In this paper we consider implications for meteoroid structure and ablation from dual peak meteor light curves, in particular looking at whether these observations support a dustball model.

 \cite{Jacchia1955} suggested that photographic observations of shortened trails, flares and meteor wake supported a dustball structure for at least some meteors.  However the fragile dustball structure proposed was inconsistent with bright meteors that survived high pressures to low observed heights. 

\cite{HawkesJones1975} developed a two component dustball model that would fit both faint and brighter meteor light curves. The key idea of this model was that meteoroids have two components, a grain component with a silicate metallic composition that is responsible for the light production, and a glue component (possibly organic in nature) which has a lower boiling point and does not produce significant luminosity.  With this model meteoroids are not necessarily fragile.  Under this model some grains will be released early (as the glue reaches its boiling point) while other grains will be separated during intensive light production by the meteor.  \cite{FisherEt2000} reviewed the observational evidence in favour of this dual component dustball model (e.g. shorter more symmetric light curves, relative independence of heights of meteors past a certain mass cutoff, good prediction of ablation profiles across a wide mass range).

\section{Observations}
The Canadian Automated Meteor Observatory (CAMO) described in detail by  \cite{WerykEt2013} uses image intensified CCD meteor observations at two stations (baseline $\approx$45~km) coupled with automatic detection and analysis software to provide a large sample of faint meteor light curves, atmospheric trajectory and orbital data.  At each CAMO  station are wide field cameras ($28^{\circ} \times 21^{\circ}$) operating at 80~fps. Each station also has a tracking high resolution system at each station with a field of view of $1.5^{\circ} \times 1.1^{\circ}$ operating at 110~fps.  The tracking system is capable of moving across the sky at 2000 deg/s, and is updated in position at 2000 Hz. Only a subset of the events are well tracked with no transverse smearing or leaving the field of view. The high resolution data will be considered in detail in a subsequent paper. Both wide and narrow field systems use Gen III image intensifiers coupled to CCD cameras with the main  spectral response from 500 to 850 nm.  

The CAMO system observes some events that have two clearly defined peaks in the meteor light curve. A typical dual peak light curve is shown in Figure~1. We have analyzed 21 double peak events for this paper. Almost all of these events had a qualitatively similar light curve structure such as that shown in Figure~1. The light curves had a rounded first peak, followed by an almost linear (on a plot using the logarithmic astronomical magnitude) increase for the second peak, followed by a rounded but usually short duration end of the second peak.
\begin{figure}
\centerline{\includegraphics[width=10.0cm]{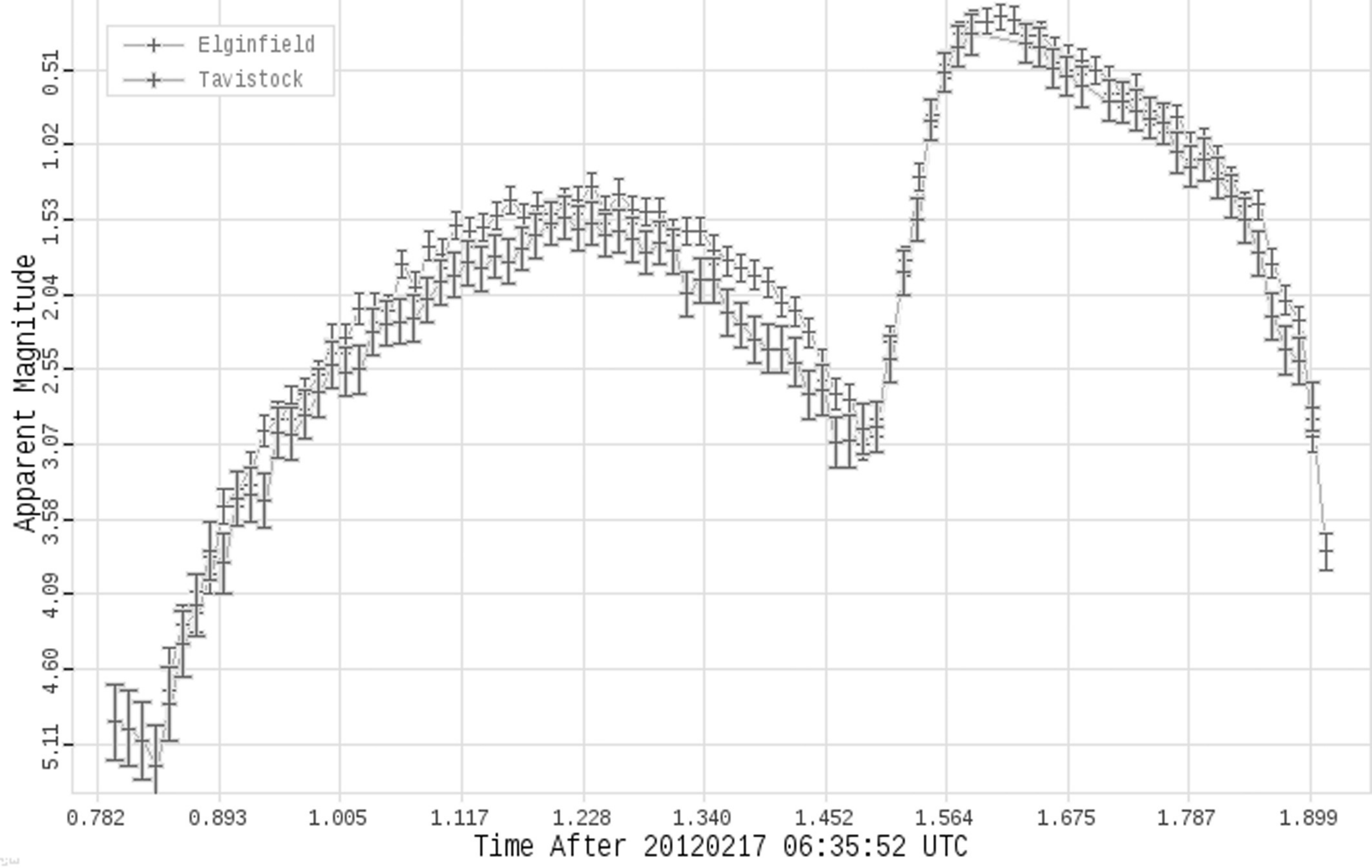}}
\footnotesize
\caption{
A plot of apparent astronomical magnitude versus time for the light curve for a meteor observed on Feb 17, 2012 at 06:35:52 UT. The dual peaks in the light curve and agreement between the observations at the two stations are clearly evident. }
\normalsize
\end{figure}

We studied the atmospheric trajectories and orbital parameters of double peaked meteor light curves, but there were no obvious differences from the general population observed with CAMO. The events studied here ranged in peak brightness between astronomical magnitude -2 and +4, and in beginning heights all but three of the events were in the range from 110 km to 80 km and none were above 120 km (although the triangulation overlap optimization may partly contribute to this). The events were somewhat slow when compared to the general population, with 42\% having speeds of 20~km/s or less. We plot in Figure~2 the Tisserand parameter plot for the double peak events which suggests that both cometary and asteroidal origin meteoroids contribute to the double peak events (generally $T_J>3$ considered asteroidal). \cite{JopekWilliams2013} discuss alternative methods to classify meteoroid orbits.
\begin{figure}
\centerline{\includegraphics[width=8.0cm]{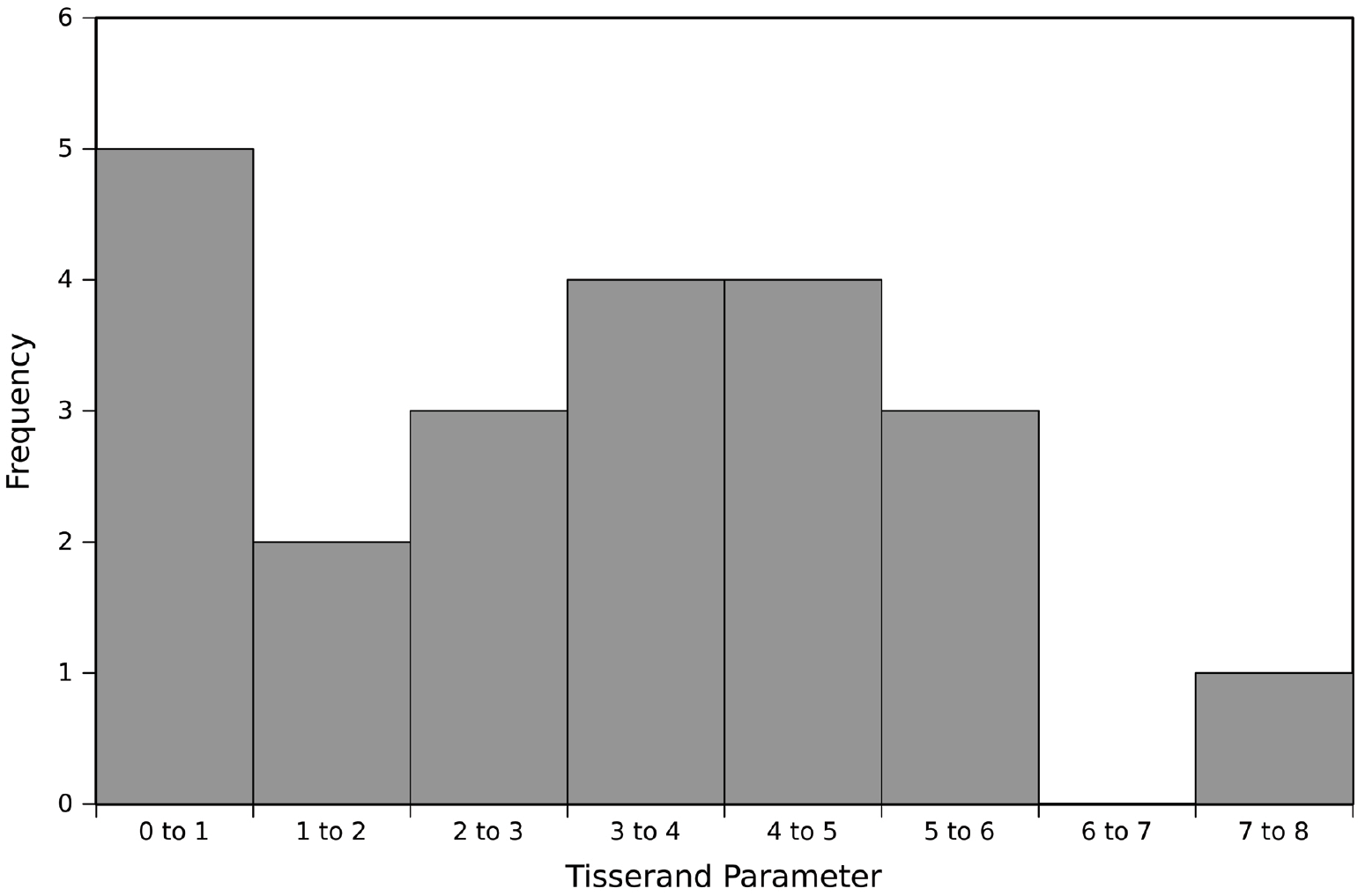}}
\footnotesize
\caption{
A plot of the distribution of the number of double peak events in this study according to the Tisserand orbital parameter. The double peak events have representation from both cometary ($T_J<3$) and asteroidal ($T_J>3$) type orbits. }
\normalsize
\end{figure}

While the events studied here were identified from the CAMO records by observers, a subsequent automated search through 198 meteors seeking events with at least a 1 magnitude difference between local minima and local maxima identified 13\% of the events as being double peaked. 
\section{Potential Meteoroid Structures}
A number of different mechanisms can produce dual peak meteor light curves.  Perhaps the most obvious explanation would be two near simultaneous parallel meteors. If the spatial resolution of the observing system is insufficient to resolve these as parallel light curves, then the result will be a dual peak meteor light curve. At least occasionally such events have been observed with high resolution systems (\cite{KaiserEt2004}). 

A second possibility is that differential chemical ablation occurs, with the peak due to the more volatile chemical species ablating first. While historical theoretical treatments considered meteoroids as chemically homogeneous, clearly actual meteoroids are collections of different chemical constituents with different thermodynamic properties.  \cite{VondrakEt2008} describe the CAMOD numerical system for predicting meteor ablation when differential ablation is taken into account. \cite{BorovickaEt2007} among others have provided spectral support for the idea of differential ablation.   
 
A third possibility is that there is a single meteoroid, but it has varying chemical or physical structure, with different layers ablating at different rates.  This is different from the second model in that different regions of the meteoroid have a different chemical composition (or possibly a different physical parameter such as density). The idea of meteoroids having coatings is broadly consistent with the literature on the structure of much smaller interstellar grains by \cite{GreenbergLi1999}.

A fourth possible mechanism is a conventional meteor ablation profile coupled with enhanced grain release either because of thermal or aerodynamic fragmentation. \cite{Simonenko1968} invoke this mechanism to explain meteor flares in shower meteors.  

A final option might be a dustball model with some grains having been released prior to atmospheric luminosity (those producing the first peak) followed by a subsequent major grain release. \cite{HawkesJones1975} saw this as a routine occurrence, although how obvious it would be, and whether it would result in two distinct peaks, will depend on the properties of the luminous grains and the glue. Of course it is quite possible that some combination of these proposed mechanisms is responsible for the light curve.
\section{Numerical Simulations}
We used a quartic Runge-Kutta approach with fixed step size to numerically model the heating and ablation of dustball grains.  The standard equations of meteor ablation were used (\cite{FisherEt2000}) and the same value for the physical and thermodynamic parameters as in that work. It was assumed that the meteoroids were sufficiently small and high enough in the atmosphere that free molecular flow dominates. The model did not incorporate sputtering as it is of negligible importance in the mass and velocity regimes considered here.  The MSISE-90 atmospheric model (\cite{Hedin1991} was used to provide atmospheric density profiles. Results from individual grains are then combined to produce a composite light curve. 

The rapid rise of the second peak in almost all of the dual peak light curves suggested to us a model with sudden release of additional grains part way through the light curve. We assumed that the second release of grains occurs at a single point.  We will refer to the grains released high as outer grains, and those released late as core grains. We show in Figure~3 one of the dual peak meteors and the fit to the light curve provided through this model. 
\begin{figure}
\centerline{\includegraphics[width=8.0cm]{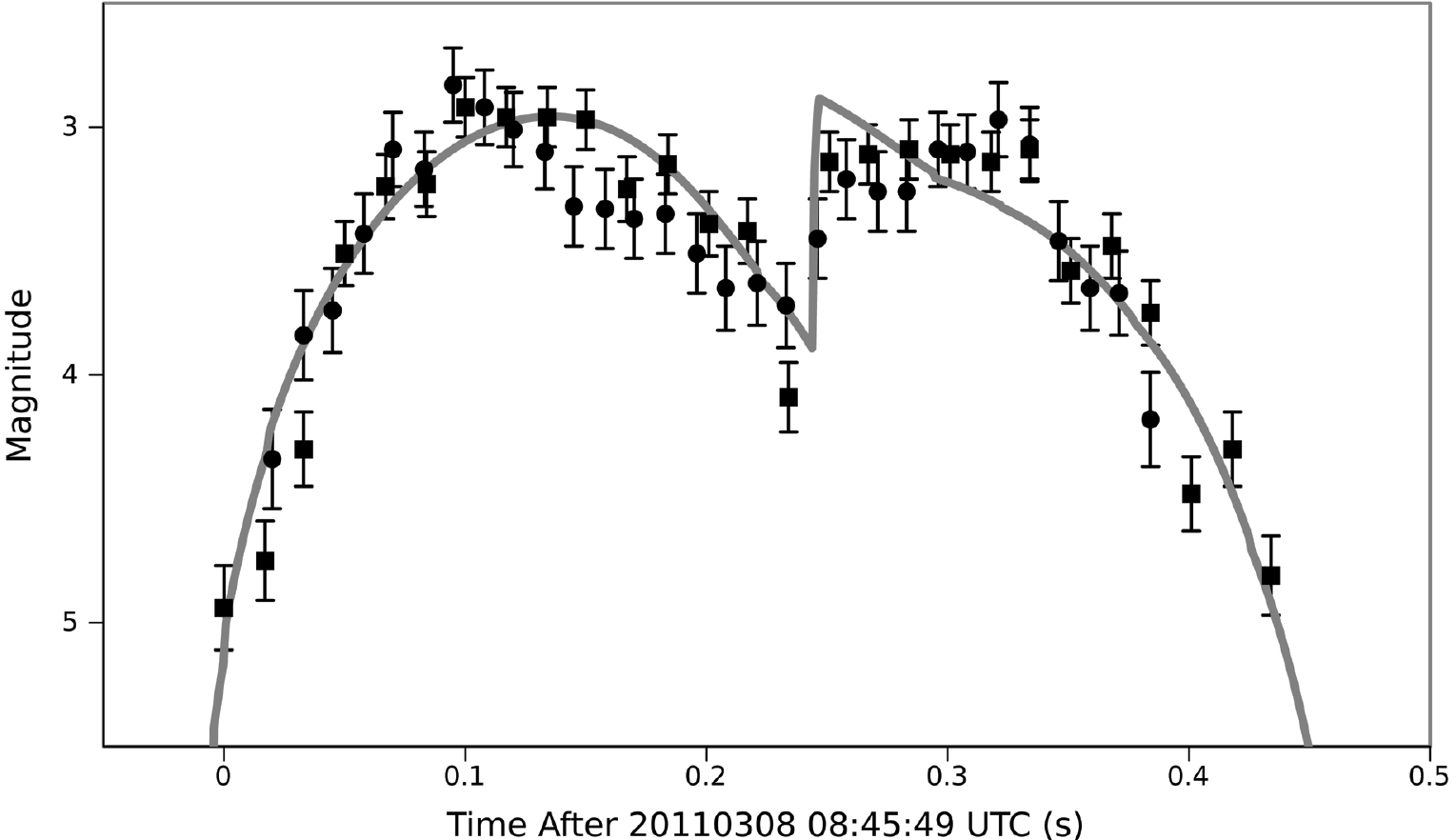}}
\footnotesize
\caption{
A plot of light curves from both wide field station cameras (blue points are from Elginfield and green from Tavistock station) along with a fit using the grain model. This meteor had a speed of 44 km/s, and the light curve extended from 103 to 86 km.  The first peak for this event was modelled with outer grains with the following characteristics: 100 $3 \times10^{-9}$ kg grains, 500 $7\times10^{-10}$ kg grains, 5000 $3\times10^{-10}$ kg grains and 50000 $2\times10^{-11}$ kg grains. The second peak was modelled with a mix of three grain sizes (all simultaneously released) 100 grains of mass  $1\times10^{-9}$ kg, 80 grains of mass  $5\times10^{-9}$ and 60 grains of mass  $2\times10^{-9}$ kg.}
\normalsize
\end{figure}

Many of the double peak light curves could be adequately matched with this simple model. While the computational model results reported here are preliminary, we have summed over all events that could be successfully modelled to obtain the mass distribution of the core and outer grains.  The data for the total mass in each grain mass range is shown in Figure~4.  An interesting result is that the core grains are larger in mass than the early release outer grains.  The core grains range from $10^{-10}$  to $10^{-5}$ kg, with a peak at $10^{-6}$~kg. The outer (pre-released) grains range from $10^{-12}$ kg  to $10^{-6}$ kg, with a peak at $10^{-9}$ kg. It should be noted that we have used the traditional luminous efficiency factor and velocity distribution (that used by \cite{FisherEt2000} and many other studies).   \cite{WerykBrown2013} have recently suggested on the basis of simultaneous radar and electro-optical observations that the video meteor mass scale is an order of magnitude smaller than previously thought. This would not change the result that core grains are larger, but would shift the actual grain sizes to a smaller size.
\begin{figure}
\centerline{\includegraphics[width=14.0cm]{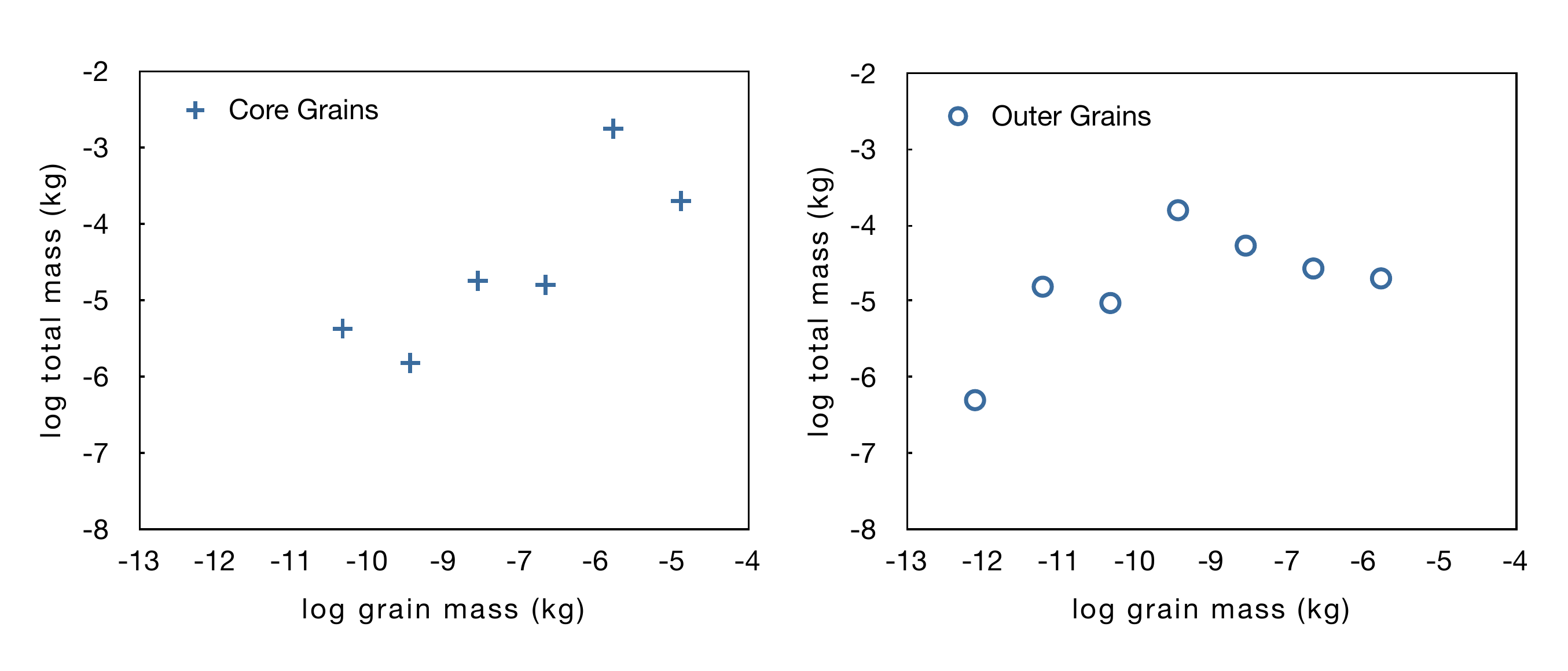}}
\footnotesize
\caption{
A plot of total mass in different grain mass categories for all events successfully modelled.  The core grains released to produce the second light curve peak are shown on the left, and the early release grains on the right. This preliminary data suggests that the outer grains are smaller in size than the core grains. }
\normalsize
\end{figure}

\section{Conclusions and Discussion}
Perhaps the most obvious question is why some meteors, but not all, show a dual peak structure.  A related question is why these events have not been more frequently reported in the past.  One possibility is that the relatively high response in the near infrared of the CAMO  detection systems may play a factor.

One of the most striking aspects of the study was the similarity of  different light curves. In almost all cases the first peak is more rounded and symmetric, while the second peak has a sharp almost linear (in a plot using the logarithmic astronomical magnitude as the vertical axis), followed by a fairly sharp rounded decline. This suggests to us that a similar mechanism is producing most of the double light curve events.  While other mechanisms can probably also match the results, the simple dustball model used here seems adequate for the events studied. \cite{MalhotraMathews2010} have conducted a statistical study of smaller meteoroids observed with large aperture radar, and find that only about one-quarter are consistent with simple single body ablation. They find that 48\% seem to show fragmentation, while 20\% imply differential ablation is important.

If a more detailed future analysis continues to suggest that the core grains are larger than the early release grains, that is an important and somewhat surprising result. This work suggests that core grains  are typically~$10^{-6}$~kg while  early release grains are of the order of ~$10^{-9}$~kg.  A number of previous studies have sought to establish the size of meteor constituent grains. \cite{Simonenko1968} found from an analysis of rapid onset flares on bright meteors a mean grain mass of about  $2\times10^{-9}$~kg.

\cite{BorovickaEt2007} have modelled six electro-optical Draconid meteors, using an erosion model in which grains are released over generally the first half of the light curve, although for the brightest meteor a number of grains were resistant to release and separated later. They generally found agreement using grains from $10^{-11}$ to $10^{-9}$ kg. They find a model with total pre-release prior to intensive ablation inconsistent with the meteors in this small sample. Of course it should be kept in mind that these were Draconid meteors, known to have a porous low density structure (\cite{BorovickaEt2007}).

\cite{CampbellKoschny2004} have modelled Leonid light curves using a thermal erosion two component dustball model. The  find that two of the meteors are well matched with Gaussian mass distribution of grains, while the third requires a power law distribution.  Grain sizes generally range from $1\times10^{-11}$ to $4\times10^{-7}$~kg, in general agreement with this study.  \cite{BeechMurray2003} used a power-law mass distribution to match Leonid light curves, generally finding the need for grains in the mass range from $10^{-10}$ to $10^{-7}$~kg. In an analysis of four Leonid fireball bursts \cite{HawkesEt2002} found that relatively large grains $10^{-5}$ to $10^{-4}$~kg were needed to match the observations.

It is also interesting to compare the grain size distribution with that obtained by Stardust during encounter with the coma of Comet 81P/Wild~2. \cite{Green2007} show that most grains are smaller than those reported here, although the total range is about $10^{-15}$ to $2\times10^{-5}$~kg.

It is possible that the outer grains released prior to ablation are produced in a similar manner to the less spatially and temporally constrained clusters of meteoroids occasionally observed (\cite{WatanabeEt2003, PiersHawkes1993}). While \cite{WatanabeEt2003} discusses possible production mechanisms, there is not yet clarity as to how these clusters occur.

\cite{BroschEt2004} point out the many parameters of meteor light curve analysis, and how they can help to constrain meteoroid structure and ablation. \cite{StokanEt2013} have studied optical trail widths using the CAMO system and this information will help constrain the possible meteoroid models.  We plan to in a later paper incorporate the tracking data for a small number of events in a more in depth investigation of dual light curve events which may help to further constrain possible models.
\begin{acknowledgments}
We acknowledge research funding from the Natural Sciences and Engineering Research Council of Canada and the Canada Foundation for Innovation. IDR received research assistantship funding from a MY Bell Research Fellowship at Mount Allison University.
\end{acknowledgments}

\end{document}